\documentclass[sigconf,authorversion]{acmart}

\usepackage{enumitem}
\usepackage{xspace}
\usepackage{amsmath,amsfonts}
\usepackage{algorithmic}
\usepackage{array}
\usepackage[caption=false,font=normalsize,labelfont=sf,textfont=sf]{subfig}
\usepackage{textcomp}
\usepackage{stfloats}
\usepackage{url}
\usepackage{verbatim}
\usepackage{graphicx}
\usepackage{balance}
\usepackage{adjustbox}
\usepackage{booktabs}
\usepackage{multirow}
\usepackage{arydshln}
\usepackage{soul}
\usepackage{fontawesome}
\usepackage{makecell}
\usepackage{natbib}
\usepackage{siunitx}
\usepackage{microtype}
\usepackage{colortbl}
\definecolor{mygray}{gray}{.9}
\usepackage{tikz}
\usepackage[T1]{fontenc}
\usepackage{aecompl}
\usepackage{xparse}
\usepackage{graphicx}

\AtBeginDocument{
  \providecommand\BibTeX{{
    \normalfont B\kern-0.5em{\scshape i\kern-0.25em b}\kern-0.8em\TeX}}}

\newcommand*\circled[1]{\tikz[baseline=(char.base)]{\node[shape=circle,draw,inner sep=1pt] (char) {#1};}}

\newcommand{\claude}{Claude 4 Sonnet\xspace}
\newcommand{\gemini}{Gemini 2.5 Pro\xspace}
\newcommand{\gpt}{GPT 4.1\xspace}

\newcommand{\Comment}[1]{}

\NewDocumentCommand{\framecolorbox}{oommm}
 {
  \IfValueTF{#1}
   {\IfValueTF{#2}
    {\fcolorbox{#3}{#4}{\makebox[#1][#2]{#5}}}
    {\fcolorbox{#3}{#4}{\makebox[#1]{#5}}}%
   }
   {\fcolorbox{#3}{#4}{#5}}%
 }

\usepackage{xcolor}

\definecolor{takeawaygray}{gray}{0.90}

\newenvironment{takeaway}[1][]{%
  \par\vspace{0.75\baselineskip}%
  \noindent
  \colorbox{takeawaygray}{%
    \parbox{\dimexpr\linewidth-2\fboxsep}{%
      \textbf{Takeaway%
        \ifx#1\empty\else\ for RQ#1\fi}%
    }%
  }%
  \par\vspace{0.5\baselineskip}%
}{%
  \par\vspace{0.5\baselineskip}%
}

\copyrightyear{2026}
\acmYear{2026}
\setcopyright{cc}
\setcctype{by}
\acmConference[ICSE '26]{2026 IEEE/ACM 48th International Conference on Software Engineering}{April 12--18, 2026}{Rio de Janeiro, Brazil}
\acmBooktitle{2026 IEEE/ACM 48th International Conference on Software Engineering (ICSE '26), April 12--18, 2026, Rio de Janeiro, Brazil}
\acmPrice{}
\acmDOI{10.1145/3744916.3773250}
\acmISBN{979-8-4007-2025-3/26/04}

\begin{document}

\title{More with Less: An Empirical Study of Turn-Control Strategies for Efficient Coding Agents}

\author{Pengfei Gao}
\affiliation{%
  \institution{ByteDance}
  \city{Beijing}
  \country{China}}
\email{gaopengfei.se@bytedance.com}

\author{Chao Peng}
\authornote{Corresponding author.}
\affiliation{%
  \institution{ByteDance}
  \city{Beijing}
  \country{China}}
\email{pengchao.x@bytedance.com}

\renewcommand{\shortauthors}{Gao et al.}



\begin{abstract}
LLM-powered coding agents, 
which operate in iterative loops (turns) to solve software engineering tasks, 
are becoming increasingly powerful. 
However, their practical deployment is hindered by significant and unpredictable costs. 
This challenge arises from a combination of factors: 
quadratically growing token counts with each turn, 
the high price of state-of-the-art models, 
the large number of turns required for real-world tasks, 
and the tendency of agents to take inefficient or unnecessary actions. 
While existing research focuses on optimizing individual turns, 
the strategic control of the total number of turns remains an underexplored area for managing agent performance and cost.
To address this gap, we conduct a comprehensive empirical study on the SWE-bench benchmark 
using three state-of-the-art models (\claude, \gemini, and \gpt). 
We systematically evaluate the impact of three distinct turn-control strategies: 
an unrestricted baseline, a fixed-turn limit with reminders, and a novel dynamic-turn strategy that grants extensions on-demand.
Our findings first reveal a fundamental trade-off in the unrestricted setting, 
where no single model excels across performance, cost, and turn efficiency. 
We then show that a fixed-turn limit, specifically at the 75th percentile of the baseline, 
serves as a "sweet spot", substantially reducing costs (by 24\%-68\%) 
with minimal impact on solve rates. 
Most significantly, our proposed dynamic-turn strategy consistently outperforms fixed-limit approaches, 
achieving comparable or better solve rates while further reducing costs by an additional 12\%-24\% 
by intelligently allocating resources only to tasks that need them. 
This work provides the first systematic analysis of turn-control strategies,
offering simple yet effective guidelines for developers to balance cost and efficacy. We demonstrate that dynamic resource allocation is a superior, easy-to-implement approach for deploying powerful yet economically viable coding agents.

\end{abstract}

\maketitle

\section{Introduction}

Large Language Models (LLMs) have profoundly impacted software development, 
with their most prominent application being the \textbf{Coding Agent}. 
A coding agent typically comprises an LLM at its core, a set of tools (e.g., a file editor, a shell), 
and an interactive environment that includes the project's codebase. 
The agent operates in a multi-round loop to solve a given task, such as a GitHub issue. 
Each cycle in this loop, referred to as a turn, involves the LLM analyzing the problem state, 
selecting a tool to interact with the environment, 
executing it, and then processing the output to decide its next action. 
This loop continues until the agent determines the task is complete. 
This powerful paradigm is already deployed in popular AI-powered IDEs like Cursor~\cite{cursor}, TRAE~\cite{traeagent2025} and CLI assistants like  Claude Code~\cite{claude_code}, and Gemini CLI~\cite{gemini_cli}.

\label{sec:introduction}
\begin{figure}
    \centering
    \includegraphics[width=.5\textwidth]{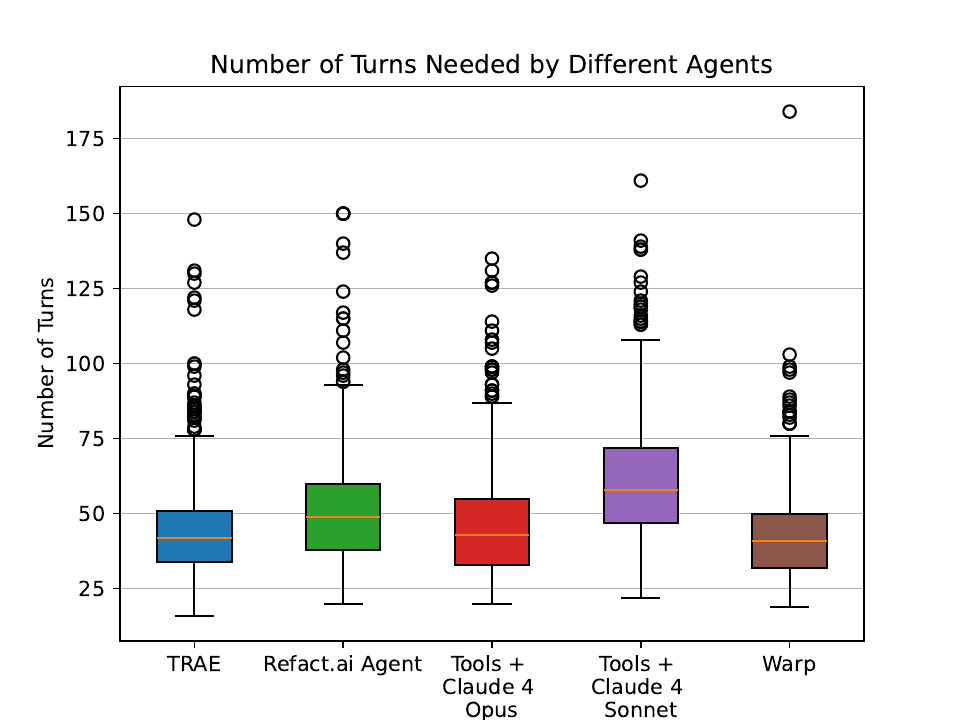}
    \caption{Number of Turns Needed by Different Agents}
    \label{fig:number_of_turns}
\end{figure}

However, a critical challenge limiting the widespread, cost-effective deployment of these agents is their significant resource consumption, 
particularly in terms of tokens. 
We argue that managing the number of turns is a crucial, yet underexplored, strategy for controlling this cost. 
Our argument is built on the following observations:
\circled{1} First, the token cost of an agentic interaction can grow faster than linearly with the number of turns. 
In every single typical agentic loop, the entire conversation history, including all previous prompts, tool calls, and their output, 
is fed back into the LLM as context for the next turn. This leads to a prompt size that grows non-linearly, often quadratically $(O(n^2))$, 
with the number of turns $n$, causing computational and financial costs to escalate rapidly.
\circled{2} Second, the state-of-the-art (SOTA) models best suited for these tasks are expensive. 
For instance, Anthropic's Claude 4 Sonnet is priced at \$3.00 per million input tokens and \$15.00 per million output tokens. 
Its more powerful counterpart, Claude 4 Opus, costs \$15.00 and \$75.00 per million input and output tokens, respectively. 
When combined with the quadratic growth of tokens, high turn counts directly translate to prohibitive costs.
\circled{3} Third, solving real-world tasks requires a large number of turns. 
An analysis of top-performing agents on the SWE-bench Verified leaderboard~\cite{swebenchleaderboard}, 
a benchmark of real GitHub issues, illustrates this reality. 
As shown in Figure~\ref{fig:number_of_turns}, 
the median number of turns required by five leading agents to solve an instance ranges from 41 to 58, 
with some complex tasks demanding over 175 turns. 
In our own experiments using \claude, 
an agent with no turn limit cost an average of \$5.85 to generate a patch and \$7.80 for a correct patch, 
underscoring the substantial expense.
\circled{4} Finally, agents are prone to wasting turns on unnecessary or even harmful actions. 
Without explicit constraints, models can be verbose or take indirect paths to a solution. 
For example, users on the Cursor IDE forum have reported agents adding "duplicate functions" and "redundant code" 
while fixing a simple bug, or breaking a project by attempting to refactor and add unrequested features~\cite{cursorforum2024}. 
These behaviors not only inflate the turn count and cost but can also degrade the quality of the final output.

Given these factors—quadratic token growth, high model prices, the high number of turns needed for real tasks, and the agent's tendency toward inefficient actions, 
we conclude that developing effective turn-control strategies is a critical and urgent research direction. 
Such strategies are of immense practical value to both end-users and the developers building these agentic products.

While research into improving the efficiency of LLM-based agents is active, 
existing work has a unifying limitation: 
it focuses on optimizing the cost or content of individual turns, 
not the number of turns in the agentic loop. 
Current strategies include optimizing the prompt via compression \cite{jiang2023llmlingua} 
and RAG \cite{lewis2020rag}; 
re-architecting the agent with plan caching \cite{zhang2025cost} 
or multi-agent systems \cite{gandhi2024budgetmlagent}; 
and constraining the output with per-turn token budgets \cite{han2024token}. 
These techniques address the payload of each step, 
not the length of the entire workflow. 
The common use of a static step limit as a safeguard \cite{jimenez2023swebench, cognition2024swebench} 
is a brute-force approach that lacks the adaptability needed for tasks of varying complexity. 
Consequently, the explicit study of dynamic turn-control strategies as a mechanism for cost-performance management in coding agents is not yet well-established.

To bridge this gap, we conduct a comprehensive empirical study to investigate the impact of various turn-control strategies on the performance and cost of coding agents. 
Our study is grounded in the realistic and challenging SWE-bench Verified benchmark, 
utilizing a representative subset of 100 tasks involving both bug fixes and feature implementations. 
We evaluate three SOTA closed-source models: \claude, \gemini, and \gpt. 
We first establish a baseline with an unlimited-turn agent. 
We then investigate a fixed-turn strategy, where agents operate under a hard deadline with explicit reminders of their remaining turns. 
Finally, we propose and evaluate a dynamic-turn strategy, which starts with a conservative turn budget and grants a one-time extension only when necessary.
In addition to these quantitative experiments, 
we present in-depth case studies to qualitatively analyze the underlying behavioral changes that explain why constrained agents can outperform their unrestricted counterparts.

Our research is guided by the following Research Questions (RQs), with key findings summarized below:

\noindent\textbf{RQ1: What are the performance and cost of coding agents in an unrestricted, unlimited-turn setting?}
We first explore the baseline performance and cost of agents operating without any turn limitations, 
where the only exit condition is the agent's own decision to terminate. 
This provides a crucial reference point for different LLMs and a baseline for our subsequent experiments.

\noindent\textit{Findings.}
Our analysis reveals a clear trade-off between problem-solving effectiveness, turn efficiency, and economic cost, 
with no single model excelling across all dimensions.
\claude emerges as the most effective model, achieving a superior solve rate of 75\%. 
This performance, however, comes at the highest financial price, costing nearly 50\% more per solution than its most economical competitor at \$7.80.
In contrast, \gpt stands out as the most cost-effective option. 
It delivers a strong 62\% solve rate, comparable to \gemini, but at a significantly lower cost of \$5.19 per solution, 
making it the optimal choice for budget-constrained scenarios.
Finally, \gemini distinguishes itself as the most turn-efficient agent, 
requiring substantially fewer conversational turns to reach solutions, 
suggesting it could be the fastest to generate patches. 
However, this efficiency is undermined by verbose, non-optional reasoning steps that inflate its output token count, 
placing its final cost between the other two models. 
This highlights a critical insight: the choice of an agent depends directly on the primary objective, 
whether it is maximizing success (\claude), minimizing cost (\gpt), or prioritizing speed of resolution (\gemini).

\noindent\textbf{RQ2: How does a fixed-turn limit with reminders affect agent performance and cost?}
Here, we investigate a strategy where a hard turn limit is imposed, set to the 25th, 50th, and 75th percentiles of the baseline turn distributions from RQ1. 
To make the agent aware of this constraint, a reminder of the remaining turns is included in each prompt (e.g., "You have X turns left to finish the task"). 
This RQ examines whether a hard deadline can maintain efficacy while reducing costs.


\noindent\textit{Findings.}
Imposing a fixed-turn limit is a highly effective cost-saving strategy, 
though its impact on performance is nuanced and reveals distinct behavioral profiles across models.
We identify a clear optimal "sweet spot": 
setting the limit at the 75th percentile of the baseline turn distribution yields substantial cost reductions, 
ranging from 24\% to 68\%, with only a minor, and sometimes positive impact, on solve rates.
The most remarkable finding involves \gemini. 
At this 75th-percentile limit, it not only saw its costs cut by a staggering 68\% but its solve rate also increased by over 3\%, 
achieving a win-win outcome. 
This suggests that a moderate sense of urgency can catalyze the model to be more efficient.
Furthermore, this strategy uncovers different robustness profiles. 
While \claude and \gpt exhibit "graceful degradation" under tighter constraints (e.g., the 25th percentile), 
\gemini displays a "threshold effect", where its performance collapses catastrophically under extreme pressure. 
These insights provide critical, practical guidance for deploying agentic systems in budget-constrained environments, 
proving that "more with less" can be achieved by setting intelligent external constraints.

\noindent\textbf{RQ3: Can a dynamic-turn growth strategy further improve the balance between performance and cost?}
In this RQ, we evaluate a dynamic strategy where an agent starts with a low turn budget (e.g., the 25th percentile limit) 
and is granted a one-time extension (e.g., to the 50th percentile limit) only if it fails to produce a patch within the initial budget. 
This RQ assesses whether a "start small, add more if needed" approach can further optimize the performance-cost trade-off.

\noindent\textit{Findings.} The dynamic-turn strategy is unequivocally superior to a fixed-budget approach. 
By starting with a lower turn limit and granting an extension only to tasks that truly need it, 
this strategy achieves comparable or even better performance than a high, fixed-turn limit, 
while consistently reducing costs by an additional 12\% to 24\%.
Our results show this advantage holds true across different configurations. 
For instance, in a strategy starting with a 25th percentile turn limit and growing to a 50th percentile limit upon initial failure, 
as well as in a higher-budget setting growing from the 50th to the 75th percentile, 
the dynamic approach was consistently superior. 
It not only lowered costs in every scenario but also, 
in some cases, improved solve rates for models like \gemini and \gpt. 
The success of this strategy stems from a core principle: efficient resource allocation. 
A fixed-budget approach wastes resources by over-provisioning for simple tasks, 
whereas the dynamic strategy intelligently reserves its budget and 
allocates the extra turns precisely to the more challenging tasks that require them. 
This "on-demand" allocation proves that a staged, targeted investment is a more powerful and economical approach than 
providing a large, one-size-fits-all budget from the outset.

Our main contributions are as follows:
\begin{itemize}[leftmargin=*]
\item We provide the first comprehensive empirical study that 
systematically investigates the impact of simple yet powerful turn-control strategies on the cost and effectiveness of coding agents, 
using SOTA LLMs on a realistic benchmark.
\item Our empirical results reveal the trade-offs of different turn-control strategies, 
identifying a practical "sweet spot" for fixed-turn limits that yields substantial cost reductions with minimal impact on 
solve rates, and uncovering distinct behavioral profiles of different models under pressure.
\item We propose and empirically validate a novel dynamic-turn allocation strategy, 
demonstrating through our experiments that it is superior to fixed-limit approaches in 
achieving high efficacy with greater economic efficiency.
\end{itemize}

The remainder of this paper is structured as follows.
Section~\ref{sec:exp_setting} details our experimental setting, 
including the benchmark used, the SOTA models evaluated, 
and the specific turn-control strategies we designed.
Section~\ref{sec:study_result} presents the detailed quantitative results of our experiments, 
structured around our three Research Questions (RQs).
Section~\ref{sec:case_study} provides a qualitative analysis through two in-depth case studies, 
exploring the behavioral mechanisms behind why constrained agents can outperform unconstrained ones.
Section~\ref{sec:threats} discusses the potential threats to validity.
Finally, Section~\ref{sec:conclusion} concludes the paper and outlines future research directions.

\section{Experimental Setting}
\label{sec:exp_setting}

This section details the experimental setup, including the task and benchmark, 
the language models employed, 
the evaluation metrics used to measure performance and cost, 
and the architecture of our autonomous agent.

\subsection{Task and Benchmark}
\begin{table}[!ht]
    \centering
    \caption{Distribution of Tasks in the SWE-bench Verified Benchmark vs. Our Subset}
    \label{tab:distribution}
    \begin{tabular}{|c|c|c|}
    \hline
        Repository & \makecell{SWE-bench Verified \\ Count} & \makecell{Our Subset \\ Count}\\ \hline
        astropy/astropy & 22 & 4 \\ 
        django/django & 231 & 47 \\ 
        matplotlib/matplotlib & 34 & 7 \\ 
        mwaskom/seaborn & 2 & 0 \\ 
        pallets/flask & 1 & 0 \\ 
        psf/requests & 8 & 2 \\ 
        pydata/xarray & 22 & 4 \\ 
        pylint-dev/pylint & 10 & 2 \\ 
        pytest-dev/pytest & 19 & 4 \\ 
        scikit-learn/scikit-learn & 32 & 6 \\ 
        sphinx-doc/sphinx & 44 & 9 \\ 
        sympy/sympy & 75 & 15 \\ \hline
        Total & 500 & 100 \\ \hline
    \end{tabular}

\end{table}

Our research targets real-world software engineering tasks extracted from GitHub issues, 
including both bug fixes and new feature implementations. 
For evaluation, we utilize the widely-used SWE-bench Verified benchmark.

To balance experimental rigor with computational and financial costs, 
we curated a representative subset of 100 tasks. 
This subset was created by randomly sampling from the full SWE-bench Verified benchmark 
while preserving the original proportional distribution of tasks across the repositories. 
Table~\ref{tab:distribution} provides a statistical overview of the task distribution in our subset compared to the full SWE-bench Verified set.

\subsection{Large Language Models}

We conducted our experiments using three state-of-the-art closed-source models to ensure our results are representative of current top-tier AI capabilities: 
\claude (Anthropic), \gemini (Google), and \gpt (OpenAI).
Our study focuses exclusively on these closed-source models because, 
at the time of our experiments, 
a discernible performance gap still existed between them and their open-source counterparts, 
particularly for complex reasoning tasks like software engineering. 
Using established, commercially available APIs also provides a stable and reproducible baseline for our experiments.

The monetary cost for each interaction was calculated based on the official pricing structures of these models at the time of our experiments. 
The specific rates are detailed below:
\begin{itemize}[leftmargin=*]
    \item \claude: \$3.00 per million input tokens and \$15.00 per million output tokens.
    \item \gemini: Employed a tiered pricing model based on the prompt's context length.
    For prompts with $\leq$ 200,000 tokens: \$1.25 per million input tokens and \$10.00 per million output tokens.
    For prompts with $>$ 200,000 tokens: \$2.50 per million input tokens and \$15.00 per million output tokens.
    \item \gpt: \$2.00 per million input tokens and \$8.00 per million output tokens.
\end{itemize}


\subsection{Evaluation Metrics}
We assess the performance and cost of each agent configuration using the following metrics:
\begin{itemize}[leftmargin=*]
    \item Solve Rate (pass@1): This metric measures the percentage of tasks for which a valid patch is successfully generated. 
A patch is considered valid if it passes the corresponding test suite provided by the SWE-bench evaluation harness.
    For each task, the agent is given up to three attempts to produce a single patch. The process stops immediately once the first patch is generated. If no patch is produced after three attempts, the instance is marked as a failure.
    \item Number of Empty Patch (\#Empty): 
    This counts the number of instances where the agent fails to generate any patch at all. 
    An experimental run for a given task can have one of three outcomes: 
    generating a valid patch (a success), 
    generating an invalid patch (a failure), 
    or generating no patch (also a failure). 
    This metric specifically tracks the last case. 
    \item Number of Total Turns (\#Total Turns): The total number of interactions with the LLM an agent consumes to solve all instances.  
    \item Token Consumption (\#Input Tokens, \#Output Tokens): The total number of input (prompt) and output (completion and reasoning) tokens consumed by the LLM.
    \item Monetary Cost (Total Cost): The financial cost incurred, calculated based on the token consumption and the respective pricing of each LLM, as detailed previously. 
\end{itemize}

\subsection{Agent Design}

\begin{figure*}
    \centering
    \includegraphics[width=1\textwidth]{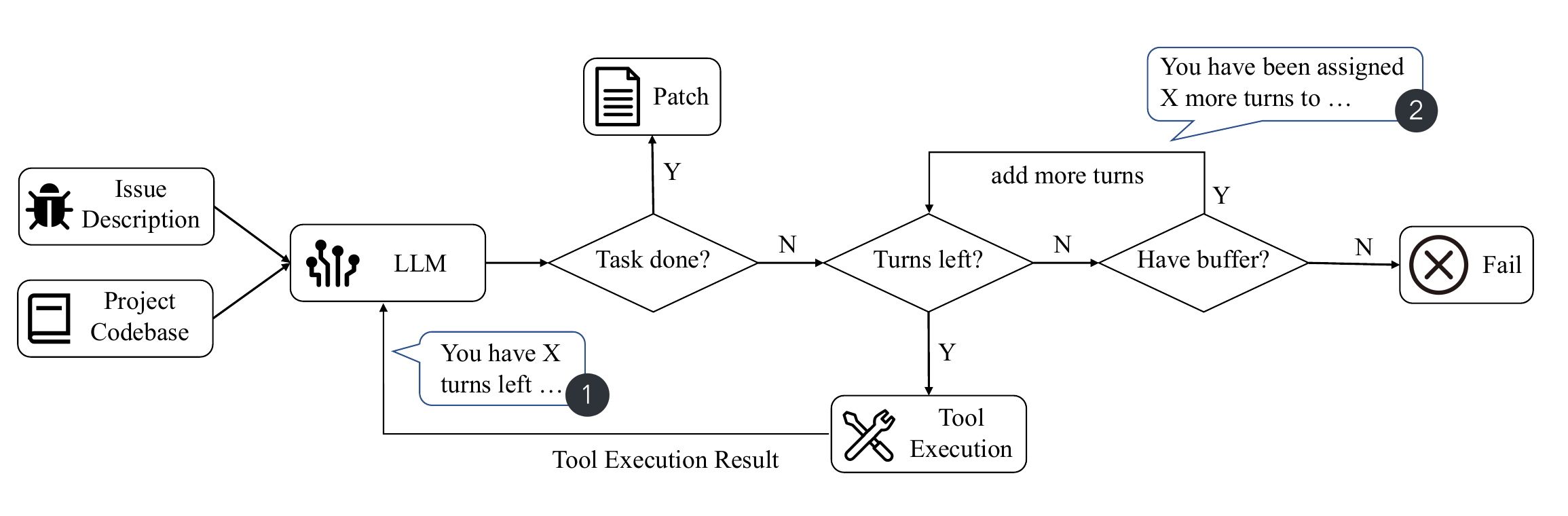}
    \caption{The workflow of the autonomous agent, illustrating the core loop of agent and the turn-control mechanism}
    \label{fig:remind_of_turns}
\end{figure*}

We adopt TRAE Agent\cite{traeagent2025}\footnote{\url{https://github.com/bytedance/trae-agent}}, the open-sourced and state-of-the-art agent on the SWE-bench Verified benchmark for our experiment. 
The agent is equipped with a specific toolset and can be deployed in one of three configurations.

\noindent \textbf{Toolset.} The agent can access to the following tools to interact with the code environment:
\begin{itemize}[leftmargin=*]
    \item \texttt{str\_replace\_editor}:  A versatile tool that allows the agent to view, create, and edit files, as well as undo previous edits.
    \item \texttt{bash}: This tool provides the agent with the ability to execute arbitrary shell commands within a secure Docker container.
    \item \texttt{task\_done}: A custom tool that the agent must call to signify that it has completed the task.
\end{itemize}


\noindent \textbf{Agent Configurations.} To investigate the effects of turn limits on task performance and cost, we tested three agent configurations.
Figure~\ref{fig:remind_of_turns} shows the overview of the agent.

\begin{itemize}[leftmargin=*]
\item \textbf{Unlimited-Turn Agent (Baseline)}: 
This agent serves as our experimental baseline. 
Its inputs are the issue description and the project codebase. 
The agent operates in a continuous loop: the LLM generates a tool call. 
If the tool is \texttt{task\_done}, 
the process terminates, and the patch is saved. 
Otherwise, the specified tool is executed, and its output is fed back to the LLM for the next turn. 
This agent has no turn restrictions.
\item \textbf{Fixed-Turn Agent with Reminder}: 
This agent is constrained by a predefined number of turns. 
To introduce a sense of urgency, at each step, 
followed by the tool execution result, a reminder prompt is added 
"ENVIRONMENT REMINDER: You have X turns left to complete the task." 
If the turn limit is reached, the task is terminated. 
This configuration allows us to observe agent performance and cost under a strict deadline.
\item \textbf{Dynamic-Turn Agent with Reminder and Growth}: 
This agent starts with a fixed initial turn limit. 
If this limit is exhausted before the task is complete, 
the agent is granted a one-time buffer of additional turns and notified with a message like 
"ENVIRONMENT REMINDER: You have used up all turns but have not yet completed the task. You are granted an additional X turns to continue and complete the task." 
The agent then continues its execution until either the task is successfully completed or the 
buffer is depleted. 
This configuration explores the benefits of granting agents more opportunities to succeed 
while still controlling resource consumption.
\end{itemize}

\section{Study Results}
\label{sec:study_result}

\subsection{RQ1: Performance and Cost on Unlimited-Turn Agent}

\begin{table*}[htbp]
\centering
\caption{Agent Performance Metrics in the Unlimited-Turn Configuration}
\label{tab:unlimited_turn}
\resizebox{\textwidth}{!}{
\begin{tabular}{lccccccccc}
\hline
LLM & Solve Rate & \#Empty & \#Total Turns & \# Turn (25th) & \# Turn (50th) & \# Turn (75th)& \#Input Tokens & \#Output Tokens & Total Cost \\
\hline
\claude & 75\% & 0 & 5,704 & 44 & 52 & 63.75 & 186,043,718 & 1,805,267 & \$585.21 \\
\gemini & 63\% & 0 & 3,925 & 19 & 28.5 & 44.5 & 239,689,532 & 4,332,151 & \$407.66 \\
\gpt & 62\% & 0 & 5,641 & 38.75 & 49.5 & 66.25 & 159,117,047 & 473,835 & \$322.02 \\
\hline
\end{tabular}}
\end{table*}

To answer RQ1, 
we conducted experiments using the unlimited-turn agent configuration. 
The comprehensive results are presented in Table~\ref{tab:unlimited_turn}. 
The first column denotes the Large Language Model (LLM) used. 
The second column indicates the percentage of problems successfully solved by the agent. 
\#Empty shows the number of instances where the agent failed to generate a patch. 
\#Total Turns represents the cumulative number of conversational turns across all problems. 
The subsequent three columns, \#Turn (25th), \#Turn (50th), and \#Turn (75th), 
detail the percentile distribution for the number of turns required to solve a single problem, 
with the median (50th) indicating typical performance. 
The \#Input Tokens and \#Output Tokens columns list the aggregate tokens consumed and generated, respectively. 
Finally, Total Cost presents the total economic expenditure in USD based on API pricing.
Our analysis is structured around three core dimensions: problem-solving effectiveness, turn efficiency, and economic cost.

\subsubsection{Problem-Solving Effectiveness}
Effectiveness is measured by the Solve Rate, which directly reflects the Agent's ability to successfully generate a valid patch. 
In this dimension, \claude demonstrated superior performance, achieving the highest solve rate of 75\%. 
This establishes it as the most effective model in an unrestricted setting. 
In contrast, \gemini and \gpt performed comparably to each other with solve rates of 63\% and 62\% respectively, 
indicating a significant performance gap when compared to \claude.

\subsubsection{Turn Efficiency}

Turn efficiency, measured by the total number of turns consumed (i.e. \#Total Turns), reflects the number of conversational steps required to finish all tasks. 
It serves as a proxy for both computational demand and time-to-solution, 
assuming consistent LLM API latency. 
As shown in Table~\ref{tab:unlimited_turn}, 
\gemini was markedly the most turn-efficient model, 
requiring only 3,925 total turns. 
Its percentile distribution further confirms that it resolved tasks more quickly on average. 
Conversely, \claude (5,704 turns) and \gpt (5,641 turns) were less efficient, 
requiring substantially more interactions to complete their tasks, with their turn counts being nearly identical.

\subsubsection{Economic Cost}
Economic cost is a critical factor for practical application, determined by token consumption and model pricing.

\gpt emerged as the most cost-effective option at \$322.02.
\claude was the most expensive model, with its total cost of \$585.21 being nearly double that of \gpt.
Interestingly, despite its superior turn efficiency, \gemini incurred high token costs (\$407.66). 
This is primarily because the model has a default "reasoning" step that generates verbose output and cannot be disabled, 
leading to a higher number of output tokens per turn.

To provide a normalized view, we calculated the average cost per successfully solved instance:
\begin{itemize}[leftmargin=*]
    \item \claude: \$7.80 per solution (\$585.21 / 75)
    \item \gemini: \$6.47 per solution (\$407.66 / 63)
    \item \gpt: \$5.19 per solution (\$322.02 / 62)
\end{itemize}
This confirms that \gpt is the most economical on a per-success basis.

\begin{takeaway}[1]
\begin{itemize}[leftmargin=*]
    \item \textbf{No One-Size-Fits-All Model:} There is no single best model; selection depends on the specific priorities of the task, balancing effectiveness, efficiency, and cost.
    \item \textbf{For Maximum Effectiveness:} \claude is the top performer, achieving the highest solve rate (75\%). This makes it the ideal choice when the primary goal is to maximize the success of problem-solving, despite its higher cost.
    \item \textbf{For Optimal Cost-Efficiency:} \gpt offers the best value. It is the most budget-friendly option and delivers a solid solve rate (62\%), matching \gemini's performance at a significantly lower cost.
\end{itemize}
\end{takeaway}

\subsection{RQ2: Performance and Cost on Fixed-Turn Agent with Reminder}
\begin{table*}[htbp]
    \centering
    \begin{tabular}{lcccccc}
    \hline
        LLM & Solve Rate & \#Empty & \#Total Turns & \#Input Tokens & \#Output Tokens & Total Cost \\ \hline
        \multicolumn{7}{c}{Unlimited Turns} \\ \hline
        \claude & 75\% & 0 & 5,704 & 186,043,718 & 1,805,267 & \$585.21 \\ 
        \gemini & 63\% & 0 & 3,925 & 239,689,532 & 4,332,151 & \$407.66 \\ 
        \gpt & 62\% & 0 & 5,641 & 159,117,047 & 473,835 & \$322.02 \\ \hline 
                \multicolumn{7}{c}{25th Percentiles of Turns} \\ \hline
        \claude & 60\% (↓ 20.00\%) & 16 & 4,107 & 98,100,806 & 1,242,319 & \$312.94 (↓ 46.53\%) \\ 
        \gemini & 26\% (↓ 58.73\%) & 63 & 1,749 & 28,026,514 & 1,613,666 & \$51.17 (↓ 87.45\%) \\ 
        \gpt & 49\% (↓ 20.97\%) & 26 & 3,415 & 59,048,594 & 238,074 & \$120.00 (↓ 62.74\%) \\ \hline
                \multicolumn{7}{c}{50th Percentiles of Turns} \\ \hline
        \claude & 63\% (↓ 16.00\%) & 11 & 4,512 & 121,324,264 & 1,350,688 & \$384.23 (↓ 34.34\%) \\ 
        \gemini & 52\% (↓ 17.46\%) & 24 & 2,314 & 51,008,477 & 1,871,124 & \$82.47 (↓ 79.77\%) \\ 
        \gpt & 57\% (↓ 8.06\%) & 10 & 4,049 & 80,391,717 & 279,172 & \$163.02 (↓ 49.38\%) \\ \hline 
                \multicolumn{7}{c}{75th Percentiles of Turns} \\ \hline
        \claude & 71\% (↓ 5.33\%) & 3 & 4,958 & 141,319,799 & 1,535,340 & \$446.99 (↓ 23.62\%) \\ 
        \gemini & 65\% (↑ 3.17\%) & 4 & 2,808 & 84,083,495 & 2,421,621 & \$129.32 (↓ 68.28\%) \\ 
        \gpt & 57\% (↓ 8.06\%) & 3 & 4,695 & 97,679,714 & 362,165 & \$198.26 (↓ 38.43\%) \\ \hline
    \end{tabular}
\caption{Performance and Cost Metrics under Fixed-Turn Limitations}
\label{tab:model_performance_limit}
\end{table*}

To address RQ2, we investigate how imposing a fixed turn limit, 
coupled with a number of turns left reminder fed to models, affects model performance and cost. 
We constrained the agents to turn limits corresponding to the 25th, 50th, and 75th percentiles of the turns consumed by their respective baseline runs (c.f. Table~\ref{tab:unlimited_turn}). 
The detailed results of this experiment are presented in Table~\ref{tab:model_performance_limit}, 
which contrasts the performance of the constrained agents against the unconstrained baseline.
The table is structured into four sections for clear comparison. 
The first section (Rows 3-5) establishes the baseline performance from the unconstrained, unlimited-turn configuration discussed in RQ1. 
The subsequent sections display the results when the turn limit is fixed at the 25th (Rows 7-9), 50th (Rows 11-13), and 75th (Rows 15-17) 
percentiles of each model's baseline turn consumption.
Within the table, the first column denotes the LLM. 
Solve Rate is the percentage of successfully solved problems. 
\#Empty counts failed attempts that yielded an empty patch. 
\#Total Turns, \#Input Tokens, and \#Output Tokens quantify resource consumption, 
and Total Cost shows the final expenditure in USD. 
Crucially, the values in parentheses () in the Solve Rate and Total Cost columns indicate the percentage change relative to the unlimited-turn baseline, 
making it easy to assess performance impact and cost reduction.

\subsubsection{Performance Impact: Convergence with a Cost}
Introducing a fixed-turn limit proved to be a double-edged sword, 
generally resulting in a degradation of problem-solving capabilities, 
albeit with a critical and insightful exception. 
The severity of the performance loss was directly correlated with the stringency of the turn limit.

Under the most restrictive 25th-percentile limit, 
all models experienced a significant decline in their solve rate. 
The performance of \claude and \gpt dropped by approximately 20\%, 
a substantial but measured decrease. 
In contrast, \gemini's performance collapsed catastrophically, 
with its solve rate plummeting by 58.73\%. 
This failure was accompanied by a dramatic increase in the number of failed patch generations (\#Empty), 
where \gemini failed in 63 instances, suggesting it is unable to finish a task under extreme pressure.

As the constraints were relaxed to the 50th and 75th percentile limits, 
performance across all models systematically improved, 
demonstrating a strong positive correlation between task success and the number of available interaction turns. 
However, a significant and counter-intuitive exception emerged with \gemini at the 75th percentile limit. 
It was the only model to not only recover but surpass its baseline performance, 
achieving a 3.17\% increase in its solve rate. 
This suggests that a moderate sense of urgency can, in some cases, 
prompt the model to generate higher-quality, more concise solutions.

This analysis also reveals distinct robustness profiles. 
\claude and \gpt exhibit a pattern of "graceful degradation", 
where their performance declines smoothly in proportion to the reduction in available turns. 
\gemini, conversely, displays a "threshold effect", 
performing exceptionally well with moderate constraints but suffering a systemic breakdown when limits become too severe.

\subsubsection{Agent Performance without Turn Reminders}

In a scenario where the agent is not reminded of the turn limit, it would simply halt upon exhausting its turns, resulting in an empty patch and a failed task. Therefore, we re-analyzed our baseline data: if an instance was successfully solved but required more turns than the given limit, we re-classified it as a failure with an empty patch. The results, summarized in Table~\ref{tab:without-reminder}, show a clear and significant performance degradation (less solve rate and more empty patches) when no reminders are given.

\begin{table*}[h]
\centering
\caption{Agent Performance with and without Turn Reminders}
\label{tab:without-reminder}
\begin{tabular}{c c c c c c}
\hline
\textbf{Percentile} & \textbf{LLM} & \textbf{Solve Rate (w/ rem.)} & \textbf{Solve Rate (w/o rem.)} & \textbf{\#Empty (w/ rem.)} & \textbf{\#Empty (w/o rem.)} \\
\hline
\multirow{3}{*}{25th}
 & Claude 4 Sonnet & 60\% & 24\% (↓ 60\%) & 16 & 73 (↑ 57) \\
 & Gemini 2.5 Pro  & 26\% & 20\% (↓ 23\%)  & 63 & 73 (↑ 10) \\
 & GPT 4.1         & 49\% & 24\% (↓ 51\%) & 26 & 71 (↑ 45) \\
\hline
\multirow{3}{*}{50th}
 & Claude 4 Sonnet & 63\% & 46\% (↓ 27\%) & 11 & 48 (↑ 37) \\
 & Gemini 2.5 Pro  & 52\% & 39\% (↓ 25\%) & 24 & 49 (↑ 25) \\
 & GPT 4.1         & 57\% & 40\% (↓ 30\%) & 10 & 48 (↑ 38) \\
\hline
\multirow{3}{*}{75th}
 & Claude 4 Sonnet & 71\% & 63\% (↓ 11\%)  & 3  & 25 (↑ 22) \\
 & Gemini 2.5 Pro  & 65\% & 54\% (↓ 17\%) & 4  & 25 (↑ 21) \\
 & GPT 4.1         & 57\% & 54\% (↓ 5\%)  & 3  & 24 (↑ 21) \\
\hline
\end{tabular}

\textit{Note:}
Rem. is the abbreviation of reminder.
\end{table*}

\subsubsection{Cost Impact: Significant and Universal Economic Benefits}
In contrast to the often-negative impact on performance, 
the fixed-turn strategy yielded universally positive and substantial economic benefits. 
Across every model and every constraint level, 
the total monetary cost was dramatically lower than the unrestricted baseline.

The magnitude of cost savings was directly proportional to the strictness of the turn limit. 
The most significant savings were observed at the 25th-percentile limit, 
where \gemini, despite its poor performance, registered a remarkable 87.45\% reduction in cost. 
Even under the most lenient 75th-percentile constraint, 
the economic advantages remained compelling, with savings ranging from 23.62\% for \claude to an impressive 68.28\% for \gemini.

\subsubsection{Synthesized Analysis: Identifying the Optimal Efficacy-Cost Ratio}
By synthesizing the performance and cost data, 
we can identify the optimal trade-off point, or "sweet spot" for this strategy. 
The 75th-percentile limit clearly emerges as the most balanced and advantageous configuration overall. 
At this level, all three LLMs achieved a highly favorable efficacy-cost ratio, 
trading single-digit performance losses (or, in \gemini's case, a gain) for cost reductions ranging from 23\% to 68\%. 
The combination of \gemini with a 75th-percentile limit was the standout performer of the experiment, 
achieving a "win-win" by simultaneously improving its solve rate and drastically cutting costs.

For applications prioritizing extreme cost-effectiveness, the performance of \gpt at the 50th-percentile limit is also highly attractive. 
It delivered a substantial 50\% cost reduction while incurring a modest performance loss of less than 10\%, 
representing an excellent value proposition.

\begin{takeaway}[2]
    \begin{itemize}[leftmargin=*]
\item \textbf{Primarily a Cost-Optimization Tool:} The fixed-turn strategy's main value lies in its ability to enforce budgetary control. It consistently yields significant cost savings, typically in exchange for a controlled degradation in solve rate.

\item \textbf{Optimal "Sweet Spot" at 75th Percentile:} Across all models, setting the turn limit to the 75th percentile of the baseline distribution provides the most favorable efficacy-cost ratio. This setting secures substantial cost reductions (ranging from 23\% to 68\%) while incurring only minimal performance losses.

\item \textbf{The "More with Less" Anomaly:} The combination of \gemini and a 75th-percentile limit is a standout exception. It not only cuts costs by over 68\% but also \textit{improves} the solve rate by 3.17\%, demonstrating that a moderate sense of urgency can catalyze a highly efficient model to unlock its potential.

\item \textbf{Actionable Recommendation:} For deploying agentic systems in budget-constrained environments, we strongly recommend implementing a fixed-turn limit. A setting at approximately the 75th percentile of the expected turn distribution is a robust strategy to dramatically reduce operational costs with negligible, and sometimes even positive, impact on performance.
\end{itemize}
\end{takeaway}




\subsection{RQ3: Performance and Cost on Dynamic-Turn Agent with Reminder and Growth}
\begin{table*}[htbp]
    \centering
    \begin{tabular}{lcccccc}
    \hline
        LLM & Solve Rate & \#Empty & \#Total Turns & \#Input Tokens & \#Output Tokens & Total Cost \\ \hline
        \multicolumn{7}{c}{Fixed 50} \\ \hline
        Claude 4 & 63 & 11 & 4,512 & 121,324,264 & 1,350,688 & \$384.23 \\
        Gemini 2.5 Pro & 52 & 24 & 2,314 & 51,008,477 & 1,871,124 & \$82.47 \\
        GPT 4.1 & 57 & 10 & 4,049 & 80,391,717 & 279,172 & \$163.02 \\ \hline
        \multicolumn{7}{c}{Dynamic 25 → 50 } \\ \hline
        Claude 4 & 64 (↑ 1.59\%) & 3 (↓ 8) & 4,150 & 101,750,021 & 1,280,320 & \$324.45 (↓ 15.56\%) \\
        Gemini 2.5 Pro & 55 (↑ 5.77\%) & 12 (↓ 40)& 2056 & 43,713,411 & 2,291,746 & \$77.56 (↓ 5.95\%) \\
        GPT 4.1 & 57 (0.00\%) & 10 (0) & 3,500 & 60,713,443 & 248,291 & \$123.41 (↓ 24.30\%) \\ \hline
        \multicolumn{7}{c}{Fixed 75} \\ \hline
        Claude 4 & 71 & 3 & 4,958 & 141,319,799 & 1,535,340 & \$446.99 \\
        Gemini 2.5 Pro & 65 & 4 & 2,808 & 84,083,495 & 2,421,621 & \$129.32 \\
        GPT 4.1 & 57 & 3 & 4,695 & 97,679,714 & 362,165 & \$198.26 \\ \hline
        \multicolumn{7}{c}{Dynamic 50 → 75} \\ \hline
        Claude 4 & 70 (↓ 1.41\%) & 1 (↓ 2)& 4,532 & 122,229,766 & 1,353,787 & \$387.00 (↓ 13.42\%) \\
        Gemini 2.5 Pro & 64 (↓ 1.54\%) & 3 (↓ 1)& 2,382 & 57,526,042 & 2,614,100 & \$98.05 (↓ 24.18\%) \\
        GPT 4.1 & 59 (↑ 3.51\%) & 1 (↓ 2)& 4,069 & 85,859,248 & 275,959 & \$173.93 (↓ 12.27\%) \\ \hline
    \end{tabular}
\caption{Performance and Cost Metrics of Dynamic vs. Fixed Turn Strategies}
\label{tab:model_performance_dynamic}
\end{table*}

RQ3 explores whether a dynamic turn-allocation strategy, 
which grants extra turns only when needed, 
can improve the balance between performance and cost. 
We define a "Dynamic X→Y" strategy as a two-stage process: 
an agent is first given a turn limit of X (e.g., the 25th percentile of baseline turns). 
If it fails to generate a patch within this initial budget, 
it receives more turns to continue its work, up to a total maximum of Y (e.g., the 50th percentile).
For instance, 
for the \claude model, 
"Dynamic 25→50" means the agent initially has a limit of 44 turns 
(the 25th percentile from Table~\ref{tab:unlimited_turn}). 
If these turns are exhausted without a patch, 
the agent is granted an additional 8 turns to complete the task, 
bringing the total maximum budget to 52 turns (the 50th percentile).

To test this approach, we conducted two comparative experiments, 
with results detailed in Table~\ref{tab:model_performance_dynamic}. 
The table is organized into two distinct comparison pairs:
\begin{itemize}[leftmargin=*]
    \item Dynamic 25→50 vs. Fixed 50: 
    The performance of the dynamic strategy (Rows 7-9) 
    is compared against a fixed 50th-percentile turn limit baseline (Rows 3-5).
    \item Dynamic 50→75 vs. Fixed 75: 
    The performance of the dynamic strategy (Rows 15-17) is compared against a fixed 75th-percentile turn limit baseline (Rows 11-13).
\end{itemize}
Within the table, the columns track key metrics such as Solve Rate, resource consumption (\#Total Turns, Token counts), 
and Total Cost. 
For the dynamic strategy results, the values in parentheses () indicate the 
percentage change relative to their corresponding fixed-turn baseline, 
directly showing performance shifts and cost savings.

\subsubsection{Analysis of the Lower-Turn Setting: Dynamic 25→50 vs. Fixed 50}

In the first comparison, 
we evaluated a dynamic-turn agent that started with 25th percentiles of turns and received a buffer to reach 50th percentiles of turns if necessary. 
The results show that this dynamic approach is clearly a better choice than simply giving the agent a fixed 50th percentiles of turns from the beginning.

In terms of performance, the dynamic strategy was either better or equal across all models. 
The solve rates for \claude and \gemini actually increased, with \gemini showing a notable improvement of nearly 6\%. 
\gpt's performance remained identical. 
This shows that giving a "second chance" to agents that struggle initially not only helps them complete their task 
but can sometimes lead to even better outcomes.

From a cost perspective, the dynamic strategy was significantly cheaper for all models. 
The cost savings were substantial, with \gpt's cost being reduced by an impressive 24.30\%. 
In short, the dynamic 25→50 strategy achieved the same or better performance than the fixed 50-turn approach, 
but at a much lower cost. Therefore, it is the unquestionably superior strategy in this setting.

\subsubsection{Analysis of the Higher-Turn Setting: Dynamic 50→75 vs. Fixed 75}
We then tested if this advantage holds in a higher-turn setting, comparing a dynamic 50→75 strategy to a fixed 75 limit. 
The findings were consistent and reinforced our earlier conclusion.

The performance of the dynamic strategy remained nearly identical to the fixed-limit approach. 
For \claude and \gemini, the performance drop was only about 1.5\%, 
a negligible difference in practice. More impressively, \gpt's resolve rate increased by 3.5\%, 
once again showing that the dynamic method can sometimes outperform a fixed, larger budget.

At the same time, the cost benefits remained significant. All models saw cost reductions between 12\% and 24\%. 
This result confirms that even when allowing for a high number of maximum turns, the dynamic strategy is a smarter and more efficient choice. 
It delivers nearly the same (or better) results while consistently saving a large portion of the budget.

\subsubsection{Discussion: The Principle of Efficient Resource Allocation}
The success of the dynamic strategy is based on a simple and powerful logic: 
most tasks do not require the maximum number of turns allowed. 
A fixed-limit strategy wastes resources by giving too many turns to "easy tasks" that can be solved quickly.

The dynamic strategy avoids this waste. It works like a "start small, add more if needed" system. 
By starting with a lower turn limit, it allows simple tasks to finish quickly and cheaply. 
It then saves its resources (the extra turns) and gives them only to the "difficult tasks" that truly need them. 
The data clearly shows that this approach of making a second, targeted investment in struggling tasks is more effective than giving all tasks a large budget from the start.

\begin{takeaway}[3]
    
\begin{itemize}[leftmargin=*]
\item \textbf{Unambiguously Superior Strategy:} The dynamic-turn strategy consistently outperforms its fixed-turn counterparts in all tested scenarios. It is demonstrably more efficient and effective.

\item \textbf{Achieves More with Less:} This approach provides a reliable method to significantly reduce costs (by 5\% to 24\%) while simultaneously preserving—and in some cases, even improving—the problem-solving success rate.

\item \textbf{Based on Efficient Resource Allocation:} Its success stems from allocating resources only when needed. It avoids the waste inherent in fixed-limit strategies by allowing easy tasks to finish cheaply and providing a "second chance" only to tasks that genuinely require it.

\item \textbf{Recommended for Practical Deployment:} For any agentic system where balancing high performance with cost management is critical, a dynamic, multi-stage turn allocation should be the default and preferred implementation strategy.
\end{itemize}
\end{takeaway}

\section{Case Study}
\label{sec:case_study}
Our quantitative results reveal a compelling, counterintuitive finding: 
agents operating under specific constraints can outperform their unrestricted counterparts. 
This section delves into the qualitative "why" behind these numbers through two illustrative case studies. 
The first case study analyzes how a moderate, fixed-turn limit can guide an agent toward a more focused and effective problem-solving strategy, 
preventing the kind of chaotic exploration that plagues unconstrained agents. 
The second case study examines the psychological and strategic benefits of a dynamic turn budget, 
showing how an initial constraint followed by a "second chance" extension can transform failure into success.

\subsection{Case Study 1}
Our quantitative results show that \gemini under a 75th-percentile fixed-turn limit with reminders, 
achieved a 65\% solve rate, 
surpassing the 63\% solve rate of its unrestricted, unlimited-turn counterpart. 
To understand the underlying behavioral changes driving this performance improvement, 
we present a case study of a instanced named sympy\_\_sympy-22456. 
This instance is particularly illustrative: the 75th-percentile agent successfully generated a correct patch in just 35 turns, 
whereas the unlimited-turn agent failed after a prolonged 82-turn trajectory that produced an incorrect patch. 
Analysis of these two trajectories reveals how moderate pressure can guide an agent toward a more effective problem-solving path.  

The successful agent, operating under the turn constraint, 
followed a direct and systematic solution path emblematic of an expert workflow. 
It correctly diagnosed the root cause of the failure: the \texttt{\_\_new\_\_} method of the String class in 
\texttt{sympy/codegen/ast.py} did not properly handle cases where an argument was already an instance of \texttt{String}, 
leading to faulty object reconstruction. 
Based on this accurate diagnosis, the agent implemented a precise and minimal fix by inserting a single conditional check, 
\texttt{if isinstance(name, cls): return name}, at the beginning of the method. 
Its verification process was equally robust; after a preliminary check, 
it correctly located the official test file, 
added a new regression test, and executed the entire module's test suite to ensure no side effects were introduced. 
The trajectory of this constrained agent demonstrates a high degree of certainty from the outset, 
suggesting that the perceived turn pressure discouraged exploratory deviations and maintained a sharp focus on a correct and efficient solution path.

In contrast, 
the unlimited-turn agent's trajectory exemplifies a cascading failure initiated by a flawed diagnosis. 
The critical divergence occurred at the initial bug-reproduction step. 
The agent received a more ambiguous error message, \texttt{TypeError: No value for 'text' given and attribute has no default}, and observed that \texttt{s.args} 
was an empty tuple. 
Instead of investigating why the arguments were empty, it incorrectly focused on the \texttt{TypeError} as the primary symptom, 
hypothesizing that the issue lay in how \texttt{s.func()} was called with no arguments. 
This misdiagnosis led it down an unproductive path of erroneous fixes. 
It spent dozens of turns repeatedly rewriting the \texttt{\_\_new\_\_} method in both the \texttt{String} class and its parents, 
introducing new \texttt{NameError} and \texttt{AttributeError} exceptions. 
This created a vicious cycle where the agent was forced to address self-inflicted errors, such as fixing a \texttt{TypeError} 
that led to a new problem with \texttt{\_\_eq\_\_}, which in turn created an issue with \texttt{\_\_hash\_\_}. 
Having lost the original objective, the agent even attempted to modify unrelated files like \texttt{cnodes.py} and became trapped in 
loops of undoing and retrying its own faulty modifications. 
The agent's reasoning became polluted by a history of failed attempts and contradictory hypotheses, making it increasingly difficult to return to a correct line of reasoning.

In summary, the failure of the unrestricted agent originated from a critical misinterpretation of an initial symptom. 
The lack of constraints afforded it the "freedom to get lost" in a protracted, 
chaotic exploration that was ultimately fatal to its success. 
The 75th-percentile agent, however, was guided by the turn limit to adopt a more focused strategy, 
leading to a correct diagnosis and an efficient, successful outcome. 
This case demonstrates that moderate constraints can act as a beneficial heuristic, 
pruning the search space and preventing the agent from pursuing erroneous, high-cost solution paths.

\subsection{Case Study 2}
The superiority of our dynamic-turn strategy is vividly illustrated in the case of 
django\_\_django-15161, 
solved by the agent with \claude. 
Under a high, fixed 75th-percentile turn limit (64 turns), 
the agent failed to produce any patch. 
However, when using a Dynamic 50→75 strategy, 
which started with a lower budget of 52 turns and received a 12-turn extension upon initial failure, 
it succeeded in generating a correct patch within the same total budget of 64 turns. 
A comparative analysis of the two trajectories reveals that the dynamic strategy's success is not merely about the total number of turns, 
but about how the phased allocation of turns influences the agent's problem-solving psychology and strategy.

The agent with a fixed 64-turn limit exhibited a behavior we term "unfocused exploration." 
Aware of its large initial budget, the agent engaged in an exhaustive but inefficient discovery process.
The agent spent its initial turns conducting a wide-ranging, unfocused exploration of the codebase. 
It viewed numerous directories and files (\texttt{django/db/models, expressions.py, \_\_init\_\_.py}), 
seemingly attempting to build a complete mental model of the entire expressions subsystem from scratch. 
This reconnaissance phase, while thorough, was not directly tied to a specific, testable hypothesis about the problem.
The agent then spent a significant number of turns (over 30) writing and repeatedly debugging a complex, 
all-encompassing test script. 
This script was designed to test the deconstruction of 15 different expression classes at once. 
This "all-or-nothing" verification strategy proved brittle; a \texttt{TypeError} in one part of the script caused repeated failures, 
forcing the agent into a protracted debugging loop that consumed the majority of its turn budget.
By the time the agent finally managed to run its tests and correctly diagnose the issue, 
it had already spent over 50 turns. 
With only a few turns remaining, 
it was unable to pivot to the implementation and final verification phase, 
ultimately timing out without producing a patch. 
The high initial budget created a false sense of security, 
encouraging a complex, non-essential verification process that ultimately led to the mission's failure.

The dynamic-turn agent, starting with a more constrained budget of 52 turns, 
adopted a markedly different and more effective strategy characterized by "focused urgency".
The initial turn limit immediately focused the agent. 
Instead of a broad file-system survey, 
it directly inspected the most relevant files (\texttt{expressions.py}, \texttt{\_\_init\_\_.py}) 
and used \texttt{grep} to quickly find all classes with decorators named \texttt{@deconstructible}. 
This was a targeted, hypothesis-driven approach aimed at understanding the specific pattern that needed to be replicated.
Rather than building a monolithic test script, 
this agent began applying the \texttt{@deconstructible} decorator to the necessary classes one by one. 
This iterative approach allowed for incremental progress and provided immediate feedback. 
The agent methodically worked through each required class (\texttt{Func}, \texttt{Value}, \texttt{ExpressionWrapper}, \texttt{When}, etc.), 
making small, verifiable changes.
The agent successfully modified all required classes just as it exhausted its initial 52 turns. 
At this critical juncture, the agent's state was highly advanced: the core logic was implemented, 
and all that remained was final verification. 
The notification that it was "granted an additional 12 turns" provided the necessary impetus to complete the task. 
The agent used this extension efficiently to run the final tests, confirm the fix, and submit the correct patch.

In summary, this case study demonstrates that the dynamic-turn strategy's effectiveness 
stems from its psychological impact on the agent's planning and execution. 
The initial, lower budget enforces focus and efficiency, compelling the agent to prioritize 
high-value actions and adopt a more direct, iterative workflow. 
The subsequent turn extension then acts as a crucial "second chance", 
providing the final resources needed to push a nearly-complete solution over the finish line. 
This contrasts sharply with the high fixed-budget approach, 
where the absence of initial pressure led to an inefficient and overly complex strategy that consumed the entire budget 
on preliminary steps. 
The total turn count was identical, 
but the phasing of the budget made all the difference, transforming failure into success.

\subsection{Stochasticity of Case Study}

We conduct supplementary repeated experiments (10 runs for each case study) to verify the stochasticity and robustness of them, as illustrated in Table~\ref{tab:repated}

\begin{table*}[]
    \centering
        \caption{Repeated Experiments for Each Case Study}
    \label{tab:repated}
\begin{tabular}{c c c c c c}
\hline
\textbf{Case Study} & \textbf{Configuration} & \textbf{Result in Paper} & \textbf{Repeated Runs} & \textbf{Success Rate} & \textbf{95\% Clopper-Pearson} \\
\hline
\multirow{2}{*}{Case 1} 
 & 75th-percentile & Solved    & 8 Solved & 80\% & [0.44, 0.97] \\
 & Unlimited-turn  & Unsolved  & 4 Solved & 40\% & [0.12, 0.74] \\
\hline
\multirow{2}{*}{Case 2} 
 & Dynamic 50->75  & Solved    & 4 Solved & 40\% & [0.12, 0.74] \\
 & Fixed 75        & Unsolved  & 2 Solved & 20\% & [0.03, 0.56] \\
\hline
\end{tabular}

\end{table*}

In two completely independent cases, our proposed strategies demonstrated a strong effect of doubling the success rate (from 40\% to 80\% in Case 1, and from 20\% to 40\% in Case 2). Due to the small sample size (n=10), a result of the high cost of these experiments, these differences do not reach statistical significance at the $p < 0.05$ level. However, both cases together reveal a clear and robust positive trend. For example, in Case 1, the lower bound of the constrained strategy's confidence interval (0.44) is already higher than the point estimate of the unlimited-turn strategy (0.40), which clearly indicates a strong trend of superior performance.

These repeated experiments greatly enhance the credibility of our case studies and the robustness of our method. They indicate that the performance improvements we observed are not accidental, but rather a systematic and repeatable positive effect of our proposed turn-control strategies.

\section{Threats to Validity}
\label{sec:threats}
\subsection{Stochasticity of LLMs}
The primary threat to validity stems from the inherent stochastic nature of Large Language Models. 
A single experimental run may not be representative of a model's average performance, 
as its output can be influenced by chance. 
A successful solution may result from a fortuitous reasoning path, 
while a failure could be an incidental anomaly. 
Although conducting our experiments on 100 tasks helps to smooth out this randomness to some extent, 
a more robust methodology would involve repeating each experiment multiple times (e.g., 3-5 runs) and 
reporting the mean and standard deviation of the metrics. 
Due to budget and time constraints, our current study did not adopt this approach, which may affect the stability of our conclusions.

\subsection{LLM Selection}

Another threat to validity lies in the selection of proprietary LLMs. Our model selection focused on state-of-the-art models from the top commercial providers, namely OpenAI, Google, and Anthropic, to ensure our findings are relevant to the most widely used, production-grade agentic systems. At the time of our experiments, on OpenRouter, the popular model service provider, nine of the top ten most-used models are closed-source models while the only open-source model is free to use, indicating that for paid workloads where cost-efficiency is critical, the market overwhelmingly relies on closed-source solutions.

The decision for Sonnet over Opus was data-driven. While Opus is a top performer, Sonnet is far more widely deployed. For instance, recent OpenRouter data shows Sonnet's minimum daily usage (33.4 billion tokens) is over five times higher than Opus's peak (6.35 billion). For a study on practical cost-efficiency, selecting the more heavily utilized model ensures our findings have a significantly greater impact on real-world agentic workloads.

\subsection{Simplified Cost Model}
Our cost calculations are based on the official token pricing of each model, 
and to ensure a consistent comparison, we uniformly disabled the prompt caching feature. 
This presents a threat to validity, 
as prompt caching is a common cost-optimization tool in real-world production environments. 
Since different models have varying pricing strategies for caching, 
enabling this feature could alter the final cost for each model, thereby affecting our assessment of their economic efficiency and the cost savings of our proposed strategies. 
Future work could investigate the cost-performance trade-offs of different turn control strategies with prompt caching enabled.

\subsection{Specificity of Agent and Prompt Design}
The conclusions of this study are contingent upon our specific agent architecture and its associated prompting strategies 
(e.g., "You have X turns left..."). 
The agent's toolset (\texttt{str\_replace\_editor}, \texttt{bash}) and the precise wording of the reminder prompts could influence its behavior. 
An agent with a more powerful toolset or different prompting techniques might respond differently to the same turn limits. 
Therefore, the observed effects may be the result of an interaction between the turn-control strategy and our specific agent implementation, 
rather than a universally applicable effect of the strategy itself.
\section{Related Work}
\label{sec:related}

In response to the escalating costs and performance challenges of LLM-based agents, 
a diverse range of optimization techniques has emerged. 
This research can be categorized by its point of intervention: 
optimizing the model's context, re-architecting the agent's structure, or constraining the reasoning process. 
A critical analysis reveals that while these methods offer value, 
they primarily address the cost and content of individual turns rather than the overall number of turns. 
This review highlights a crucial gap in the literature: the absence of dynamic, intelligent control over the length of the agent's iterative loop.

\subsection{Optimizing the Prompt}
The most direct approach to reducing cost is optimizing the prompt. 
Prompt compression methods shorten the input while preserving key semantic content \cite{hu2025dynamic, jiang2023llmlingua}. 
These generally fall into two categories: extractive techniques that prune redundant tokens based on information-theoretic metrics, 
and abstractive techniques that use a smaller model to generate a summary \cite{li2024prompt}. 
More advanced methods leverage the model's internal states \cite{fei2025efficient} 
or reinforcement learning \cite{hu2025dynamic, jiang2023llmlingua} to select the most salient information.

A more holistic approach is Retrieval-Augmented Generation (RAG), 
a form of context engineering where an agent retrieves relevant information from an external knowledge base (e.g., documentation or code) 
to augment its prompt before generation \cite{shinn2023reflexion, lewis2020rag, github2024rag, su2024evor}. 
By improving the prompt's signal-to-noise ratio, 
RAG directly addresses the "lost in the middle" problem, where models struggle to utilize information buried in long contexts \cite{liu2023lost}.

Despite their effectiveness, these techniques operate within a turn. They optimize the payload of each conversational step, 
making a 100-turn interaction cheaper but doing nothing to prevent the agent from taking 100 turns in the first place. 
Control over the iteration count remains outside their scope.

\subsection{Re-architecting the Agent}

A second category of optimizations moves beyond the prompt to re-design the agent's fundamental architecture.
Agentic Plan Caching stores and reuses structured plan templates for semantically similar tasks. 
This allows the system to bypass the expensive, multi-turn planning process for known problems, 
invoking a full planner LLM only for novel tasks. 
This has been shown to reduce serving costs by 46.62\% while maintaining 96.67\% of optimal performance \cite{zhang2025cost}.

Codified Prompting, as seen in the CodeAgents framework \cite{yang2025codeagents}, 
replaces verbose natural language communication between agent components with a structured, typed pseudocode. 
This makes the agent's reasoning more precise and token-efficient, reducing input tokens by 55\%–87\% and 
output tokens by 41\%–70\%.

Hierarchical and Multi-Agent Systems decompose complex tasks, 
delegating sub-tasks to multiple, specialized agents \cite{wu2024autogen, crewai2024, langchain2024memory}. 
This enables strategies like cost-effective cascades, 
where cheaper models handle most work and expensive models are reserved for critical steps, 
achieving cost reductions up to 94.2\% \cite{gandhi2024budgetmlagent}.

These architectural patterns apply classic software engineering principles to AI systems. 
However, they either bypass the iterative loop (caching), 
make communication within it cheaper (codified prompting), 
or delegate work from it (hierarchical systems). 
None introduce a mechanism to actively govern the iterative process of a single worker agent. 
The problem of an agent getting stuck in a loop is simply moved, not solved.

\subsection{Constraining the Reasoning Process}

The research most adjacent to our work involves direct constraints on the agent's reasoning process. 
Token Budgeting explicitly instructs the LLM to limit the length of its generated response in a given turn. 
The TALE framework, for example, dynamically estimates a "reasonable" token budget and injects it into the prompt, 
reducing token costs by an average of 67\% \cite{han2024token}. 
Other work has explored fine-tuning models to self-estimate budgets \cite{li2025selfbudgeter}. 
A key finding from this research is the "Token Elasticity" phenomenon: an overly aggressive budget can backfire, 
causing the model to generate an even longer response \cite{han2024token}.

In practice, many systems also use a Static Step/Turn Limit (e.g., max\_turns) 
as a brute-force safeguard against unlimited loops, as seen in frameworks like SWE-bench \cite{jimenez2023swebench, 
cognition2024swebench}. 
However, this is not an intelligent strategy, 
as a fixed limit lacks the adaptability for tasks of varying complexity.

The distinction between these methods and our work is critical. 
Token budgeting controls the verbosity of a single turn. 
It does not control the number of turns in the agentic loop. 
It addresses the length of an atomic unit of work, whereas the problem at hand is the length of the entire workflow.
\section{Conclusion}
\label{sec:conclusion}
The high operational cost of LLM-based coding agents poses a significant barrier to their widespread adoption. 
This paper addresses this challenge through the first comprehensive empirical study of turn-control strategies 
as a primary mechanism for cost management. 
Our findings demonstrate that well-calibrated fixed-turn limits can drastically reduce costs—by up to 68\%—with minimal impact on solve rates, 
and can even improve performance by inducing more focused agent behavior. 
Furthermore, we propose and validate a novel dynamic-turn strategy that outperforms the best fixed-limit approaches by intelligently allocating turns 
only to the tasks that require them. 
Overall, our work provides concrete, actionable evidence that explicit turn management 
is a critical lever for optimizing the cost-effectiveness of coding agents, 
offering practical guidance for building the next generation of economically viable AI developers.
In the future, we want to develop 
more sophisticated strategies where agents can dynamically request or budget turns based on a task's perceived complexity or their own solution confidence.

\section*{Data availability}
The code and related experimental data in this paper are accessible on this page: \url{https://doi.org/10.6084/m9.figshare.30705524}


\newpage

\balance
\bibliographystyle{ACM-Reference-Format}
\bibliography{reference}

\end{document}